# Suppressing carrier density in $(Bi_xSb_{1-x})_2Te_3$ films using $Cr_2O_3$ interfacial layers

Xiong Yao[1], Hee Taek Yi[2], Deepti Jain[2] and Seongshik Oh[1]

[1] Center for Quantum Materials Synthesis and Department of Physics & Astronomy, Rutgers, The State University of New Jersey, Piscataway, New Jersey 08854, USA
[2] Department of Physics & Astronomy, Rutgers, The State University of New Jersey, Piscataway, New Jersey 08854, USA

E-mail: ohsean@physics.rutgers.edu

## Abstract

Band structure engineering and interfacial buffer layers have been demonstrated as effective means to tune the Fermi level in topological insulators (TI). In particular, the charge compensated compound $(Bi_xSb_{1-x})_2Te_3$ (BST) plays a critical role in the molecular beam epitaxy growth of magnetic TIs. Here we introduce a strategy of exploiting epitaxial $Cr_2O_3$ as a buffer layer and amorphous $Cr_2O_3$ as a capping layer in the growth of BST films. These films exhibit carrier density lower than $10^{12}/cm^2$ over a wide range of Bi contents and higher mobility than BST films directly grown on $Al_2O_3$ substrate, shedding light on the importance of interfacial layers for TI films and paving a new avenue to the application of magnetic BST films.

## 1. Introduction

Topological insulators (TIs), especially when incorporated with magnetism through either magnetic doping or magnetic proximity coupling [1-7], have given rise to several discoveries in the past decade. The realization of these phenomena as well as development of topological devices requires ultralow carrier densities or extremely low temperatures to freeze the disorder effect [1-3, 8-13]. For example, tuning the Fermi level to the magnetic exchange gap is a prerequisite for the emergence of quantum anomalous Hall effect (QAHE) and axion insulator state [1-5]. To achieve this goal in TIs, a band structure engineering strategy of mixing electron-type $Bi_2Te_3$ and hole-type $Sb_2Te_3$ has been developed to realize charge compensated TI [14]. The low carrier density, strong spin-orbit coupling, and more importantly, exposed Dirac point of $(Bi_xSb_{1-x})_2Te_3$ (BST) makes it an ideal and so far the only parent material that can be magnetically doped and lead to QAHE and axion insulator state by molecular beam epitaxy (MBE).

On the other hand, it has been demonstrated that interfacial buffer layers can effectively suppress the defect density in TIs [10, 11, 15, 16]. Considering the central role of BST in magnetically doped TIs, exploiting interfacial buffer layers in BST thin films could further suppress the defect density and push the Fermi level closer to the Dirac point. Protective capping layers have been also widely used in TIs to minimize sample degradations [17, 18]. Considering that chromium is the most widely used magnetic element for both QAHE and axion insulators on magnetic BST layers, $Cr_2O_3$ can be a promising option for both buffer and capping layers for the magnetic BST system. Nonetheless, except for a couple of studies along the direction of exchange effect [19, 20], it remains unknown how effective $Cr_2O_3$ can be as a buffer for suppressing the residual carrier densities of the BST system. Here we demonstrate that by simultaneously using epitaxial $Cr_2O_3$ as a buffer layer and in-situ grown amorphous $Cr_2O_3$ as a capping layer, the carrier density can be effectively suppressed to $8.3 \times 10^{11}/cm^2$ in p-type BST sample and $9.8 \times 10^{11}/cm^2$ in n-type BST sample, so far the lowest sheet carrier densities ever reported for the BST system, suggesting that $Cr_2O_3$ is a promising buffer and capping layer for the BST magnetic TIs.

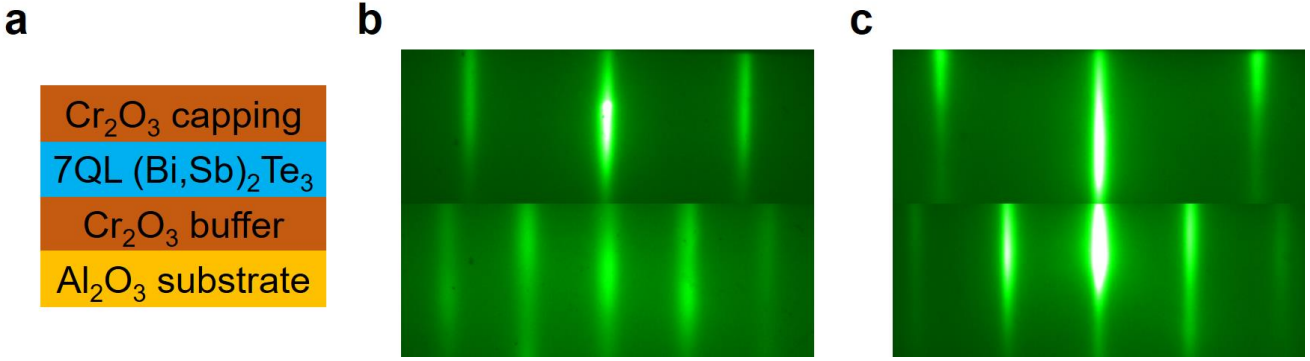

**Figure 1.** (a) Schematic of the sample structure. (b) Reflection high-energy electron diffraction (RHEED) patterns of the epitaxial $Cr_2O_3$ buffer layer grown on $Al_2O_3$ substrate, taken from two high-symmetry directions. (c) RHEED patterns of the BST layer are taken from two high-symmetry directions.

## 2. Results and discussion

We grew all the samples on 10 mm × 10 mm $Al_2O_3$ (0001) substrates using a custom-built SVTA MOS-V-2 MBE system with a base pressure of low $10^{-10}$ Torr. The sample structure is shown in Figure 1a. All the substrates were cleaned ex-situ by 5 minutes exposure to UV-generated ozone and in situ by heating to 750 ºC under an oxygen pressure of $1 \times 10^{-6}$ Torr for ten minutes. The epitaxial $Cr_2O_3$ buffer layer with thickness of 4 nm was grown by supplying Cr flux in $2 \times 10^{-6}$ Torr oxygen. The sample was then cooled down to 260 ºC and the BST layer was grown with a 10:1 flux ratio of Te to (Bi, Sb). Then, the sample was cooled to room temperature and the 4 nm amorphous $Cr_2O_3$ capping layer was grown by supplying Cr flux in $2 \times 10^{-6}$ Torr oxygen. All the source fluxes were calibrated in situ by quartz crystal micro-balance (QCM) and ex-situ by Rutherford backscattering spectroscopy (RBS).

The growth quality was in situ monitored by reflection high-energy electron diffraction (RHEED). The RHEED images of the $Cr_2O_3$ layer and BST layer are shown in Figure 1b and c. Two types of high symmetry RHEED patterns with the line spacing differing by $\sqrt{3}:1$ ratio were observed for both $Cr_2O_3$ and BST layers, consistent with their six-fold in-plane symmetries. All these patterns repeat every 60˚, confirming the ordered single domain nature of these films. All these RHEED patterns show sharp bright streaky features, indicating high-quality epitaxial growth.

One immediate advantage of using epitaxial $Cr_2O_3$ as the buffer layer (particularly on $Al_2O_3$ substrate) is creating a more active surface for the deposition of TIs. Even if $Al_2O_3$ (0001) is one of the popular substrates for TI film growth, its surface is so inert that Bi and Sb have a hard time sticking to the surface at optimal growth temperatures. In the case of $Bi_2Se_3$ films, this problem is resolved using the two-temperature-step growth scheme, developed by Bansal et. al. about a decade ago [21, 22], where the low-temperature step helps Bi atom stick to $Al_2O_3$ while extra Se desorbs away. However, this two-step growth scheme is not so effective for the BST films because, at a substrate temperature high enough to evaporate away extra Te, Bi and Sb have a hard time sticking to the substrate. It turns out that the $Cr_2O_3$ surface, being less inert than the $Al_2O_3$ surface, allows Bi and Sb to stick easily at the temperature required to make any extra Te evaporate away from the substrate.

To investigate the change of carrier density upon varying Bi content and achieve the charge neutral point (CNP), we grew five $(Bi_xSb_{1-x})_2Te_3$ samples with nominal Bi contents x of 0.2, 0.22, 0.25, 0.27, and 0.3. The transport properties of these films, including longitudinal resistance and Hall resistance, were measured in the standard van der Pauw geometry, by manually pressing four indium wires on the corners of each sample. A closed-cycle cryostat with a base temperature of 6.7 K and a magnetic field up to 0.6 T was used for the transport measurements. Raw data of $R_{xx}$ and $R_{xy}$ were properly symmetrized and anti-symmetrized, respectively.

Figure 2a gives the temperature-dependent 2D sheet resistance of these samples. All five samples display increasing resistance with decreasing temperature, reflecting depleted carriers due to the compensation effect between n-type $Bi_2Te_3$ and p-type $Sb_2Te_3$. The sheet resistance rises from x = 0.2 and reaches a maximum at x = 0.25 then gradually drops with further increasing Bi content, as shown in Figure 3a, which is consistent with the observations in BST directly grown on $Al_2O_3$ [14]. The maximal point of $R_{xx}$ at x = 0.25 is a signature of CNP. Figure 2b shows the corresponding Hall resistance measurements of these samples, which clearly demonstrate a transition from p-type to n-type carrier at x = 0.25, confirming x = 0.25 is the CNP. The Hall resistance of sample x = 0.25 exhibits nonlinear behavior while all the other samples show well-defined linear n- or p-type Hall resistance. The nonlinear Hall resistance in sample x = 0.25 is a typical feature of TIs that are in the n-p carrier mixed regime with the Fermi level near CNP [11, 16].

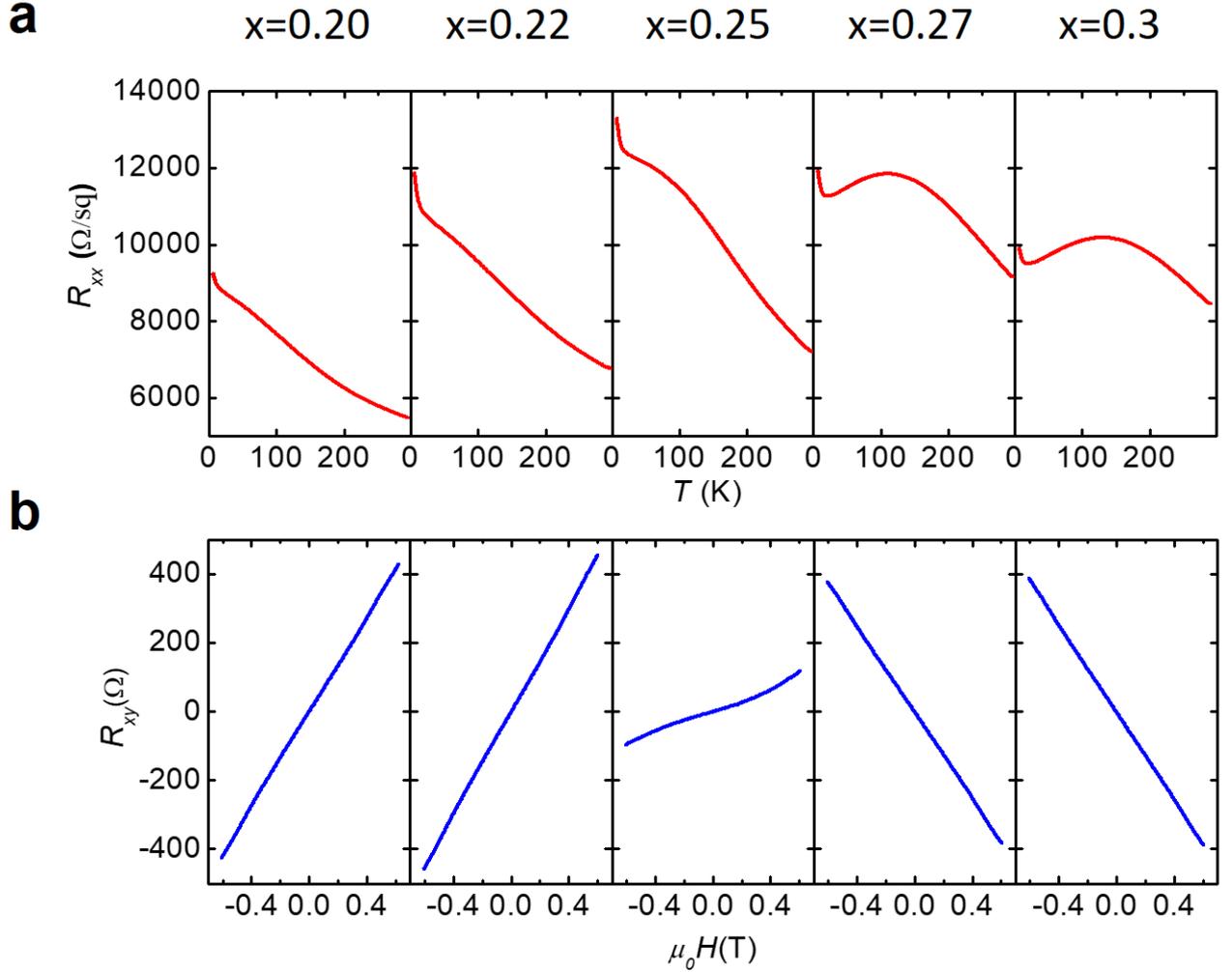

**Figure 2.** (a) Temperature-dependent longitudinal sheet resistance $R_{xx}$ measured from 300 K to 6.7 K. (b) Field-dependent Hall resistance $R_{xy}$ measured at 6.7 K and up to 0.6 T.

Figure 3b gives the 2D carrier densities of all the well-defined n- or p-type samples based on the Hall resistance. Sample x = 0.25 is not included here due to its ill-defined carrier density from the nonlinear Hall resistance. The minimal carrier density is $8.3\times10^{11}$/cm$^2$ in the p-type samples and $9.8\times10^{11}$/cm$^2$ in the n-type samples, both values being clearly lower than the BST films grown on $Al_2O_3$ and even lower than the gated BST grown on GaAs [23]. It is also notable that for all the samples in Figure 3b with Bi contents from 0.2 to 0.3, the 2D carrier densities are lower than BST/$Al_2O_3$. For comparison, the minimal carrier density in BST/$Al_2O_3$ can only be realized in a very narrow Bi content window and the carrier density quickly rises when the Bi doping level is out of this window. The low carrier density in these samples observed over a relatively wide range of Bi doping content is another merit of this system.

Combining the $R_{xx}$ and carrier density data gives the Hall mobilities of these samples, as summarized in Figure 3c. In contrast to the Bi content dependency of mobility in BST/$Al_2O_3$, which reaches a maximum near the CNP, the Hall mobility in our samples tends to decrease, while approaching the CNP. Nonetheless, all the mobility values shown in Figure 3c are still higher than the maximum value of the BST/$Al_2O_3$ system. This indicates that the interfacial $Cr_2O_3$ buffer layer helps suppress the net effects of the interfacial defects.

Further studies are required to fully explain the supressed carrier density and enhanced mobility in our samples. However, compared to the mechanically polished $Al_2O_3$ substrate, the epitaxially grown $Cr_2O_3$ layer is likely to provide more defect-free interfaces, and compared to other substrates, the matching valence states of both cations and anions to those of BST would also help reducing interfacial charge defects. Furthermore, a-Cr$_2$O$_3$ capping layer provides effective protection against environmental doping and aging effects.

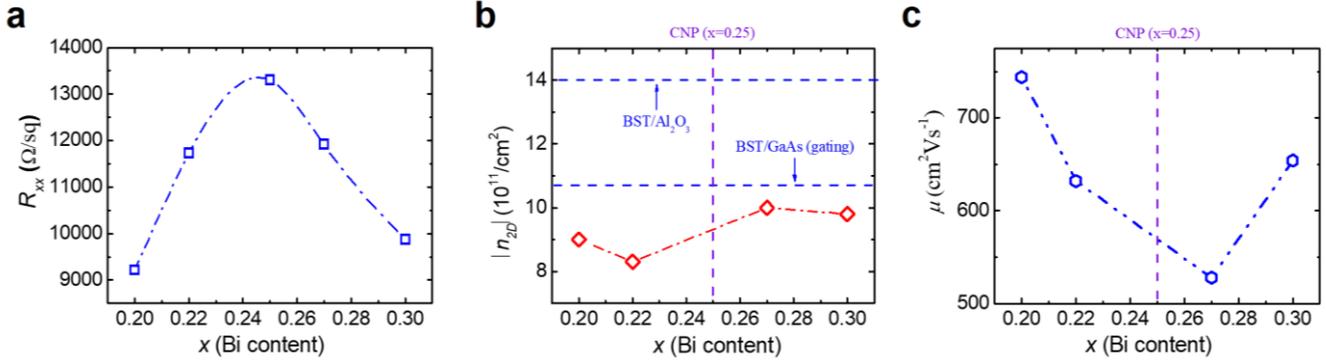

**Figure 3.** (a) Summary for the maximal values of longitudinal sheet resistance $R_{xx}$ in samples with varying Bi contents. (b) 2D carrier densities were obtained from the Hall resistance for all the samples except x = 0.25, which is not shown here due to ill-defined carrier density. Blue dash lines indicate the carrier density values of BST/Al$_2$O$_3$ from Ref. [14] and gated BST/GaAs from Ref. [23]. (c) Summary for Hall mobilities obtained from $R_{xx}$ and $R_{xy}$ data.

## 3. Conclusion

As the parent compound of quantum anomalous Hall and axion insulators, BST films play a central role in the MBE growth of magnetic TIs. By introducing epitaxial Cr$_2$O$_3$ as a buffer and amorphous Cr$_2$O$_3$ as a capping layer, BST films exhibit lower carrier densities and higher Hall mobilities than BST/Al$_2$O$_3$ over a wide doping range. Accordingly, our results demonstrate that Cr$_2$O$_3$ could work as an effective buffer and capping layer for magnetic (particularly, Cr-doped) BST films toward better quantum anomalous Hall and axion insulators.

## Acknowledgments

This work is supported by National Science Foundation's DMR2004125, MURI W911NF2020166, and the center for Quantum Materials Synthesis (cQMS), funded by the Gordon and Betty Moore Foundation's EPiQS initiative through grant GBMF6402.